%% file: paper.tex
\documentstyle{article}
\input mathmac.tex
\input webmac.tex

\input refdef.tex
\input command.tex

\begin{document}
\renewcommand{\floatpagefraction}{0.8}
\begin{titlepage}
\begin{center}
\thispagestyle{empty}
\vskip 1.5cm
\begin{flushright}
	 {\vspace*{-1.5in}
	 Fermilab--PUB--96/044--T\\
	 CERN-TH/96--50\\
         RAL-TR-96-012\\}
\end{flushright}
\vskip 2 cm
{\Large \bf The evolution of parton distributions beyond leading order:\\}
\vskip 0.2 cm
{\large \bf the singlet case\\}
\vskip 0.7 cm
{\large \bf R. K. Ellis}
\vskip 0.3 cm
Fermi National Accelerator Laboratory\footnote{Permanent address},\\
P. O. Box 500,\\
Batavia, IL 60510, USA.\\
\vskip 0.3 cm
and \\
\vskip 0.3 cm
Division TH,\\
CERN,\\
1211 Geneva 23, Switzerland.\\
\vskip 0.7 cm
{\large \bf W. Vogelsang}
\vskip 0.3 cm
Rutherford Appleton Laboratory, \\
Chilton, DIDCOT,\\
Oxon OX11 0QX, England. \\
\vskip 2 cm
\today \\
\vskip 2.5 cm
{\bf Abstract}\\
\end{center}
\vskip 0.5 cm
A complete description of the calculation of 
anomalous dimensions (GLAP splitting functions) is given 
for parton distributions which appear in space-like processes.
The calculation is performed in the light-cone gauge.
The results are in agreement with the previous results of Furmanski and 
Petronzio. 
\end{titlepage}
\section{Introduction}
The light-cone gauge has always occupied at special role in the description 
of hard processes at high energy. It belongs to a class of physical gauges 
in which many of the precepts of the QCD parton model are true, because
in this gauge collinear divergences occur in diagrams corresponding to the 
parton cascade. We thus retain the probabilistic interpretation of 
a hard scattering event, which is obscured in covariant gauges. In fact, 
by introducing an additional gauge vector we obtain many of the advantages 
of an infinite momentum frame formulation in a covariant notation.

However, doubts have been raised about the utility of light-cone gauge
in practical calculations[\ref{Leib},\ref{AT}]. One refutation of
these misgivings is provided by the classic calculation of the two
loop splitting functions or anomalous dimensions as given by Curci,
Furmanski and Petronzio[\ref{CFP}] for the non-singlet case and the
calculation of Furmanski and Petronzio[\ref{FP}] for the singlet case.
The calculation of ref.~[\ref{FP}], however, has never been fully
documented. It could be be that this lack of complete documentation
has acted as a barrier to further developments along this line.  One
example of a further application of this method is the calculation of
the polarized two loop splitting functions, recently presented in
ref.~[\ref{Vogelsang}].  It therefore seemed a valuable addition to
the literature to provide a more complete description of the
calculation of the singlet evolution probabilities.  This is the
modest aim of this paper.
 
In our calculation $n$ is a light-like vector which serves to define the
longitudinal direction. The momentum of the incoming parton 
(taken to be massless) is denoted by $p$.
Thus we have two light-like vectors which are defined such that,
\beq
n^2=p^2=0, \;\; 
n \cdot p \equiv pn \neq 0, \;\; n \cdot t=p \cdot t=0 \; ,
\eeq
where $t$ is any vector in the transverse plane.
In addition to specifying the longitudinal direction 
we also use the vector $n$ to fix the light-cone gauge:
\beq
n \cdot A=0 \; .
\eeq
Following reference [\ref{CFP}] we use the principal value (PV) prescription 
to regulate the divergences which occur in the light-cone gauge propagator 
in loop and phase space integrals, i.e.
\beq \label{PPprescription}
\frac{1}{n\cdot k} \rightarrow \frac{1}{2} \Bigg(
\frac{1}{n\cdot k+ i \delta (pn)} + 
\frac{1}{n\cdot k- i \delta (pn)} \Bigg)
=\frac{n \cdot k}{(n\cdot k)^2 + \delta^2 (pn)^2} \; .
\eeq 

This prescription is at variance with the Mandelstam-Leibrandt (ML)
treatment of the $1/(n \cdot k)$
singularities[\ref{Mandel},\ref{Leib}].  The ML prescription, since it
permits the Wick rotation in virtual diagrams, leads to power counting
rules and a proof of renormalizability of the theory in this gauge.
Indeed the one loop non-singlet splitting function has been
investigated in this gauge[\ref{Bass}].  We have chosen not to follow
the ML prescription.  The technical reason is that it leads to a
proliferation of graphs, because of ghost-like contributions
associated with the $n\cdot k$ propagator[\ref{Bass}].  The physical
reason is that we wanted to stay as close as possible to old-fashioned
perturbation theory in which manifest Lorentz covariance is sacrificed
in order to have a simple form for unitarity.  The advantage of
Eq.~(\ref{PPprescription}) is that the unitarity of the theory is
explicit.  In this way we hoped to gain a greater physical
understanding of the two loop anomalous dimensions.  It would be
interesting to repeat the calculation of the two loop anomalous
dimensions using the ML prescription.
 
Of course, the calculation of any gauge invariant quantity such as the
two loop splitting function is independent of the gauge in which the
calculation is performed and of the method of calculation. However, the
discovery of complications in the covariant gauge
calculations[\ref{HamVN},\ref{ColSc}] makes the method outlined in
ref.~[\ref{EGMPR}] and implemented in refs.~[\ref{CFP},\ref{FP}] even more 
attractive.
Not only is this method close to the parton model, but it also leads
to compact answers and may be the most efficient method from a
calculational point of view. It might present a viable 
method for the analytic calculation of the three loop 
splitting functions\footnote{Results for low moments of the three loop
splitting functions are given in Ref.~[\ref{Larin}]}.
\def\Proj{{\cal P}}
\section{Calculation of anomalous dimensions} 
\subsection{Factorization}
In this section we shall explain the method of factorization of the 
two-particle irreducible (2PI) diagram in the
light-cone gauge. It is not our intention to repeat the discussion
which is clearly provided in ref.~[\ref{CFP}]. We only include
those details which are necessary to present the structure of the
calculation or to define the notation. Following ref.~[\ref{EGMPR}]   
we define a generalized ladder expansion by introducing the 2PI 
kernel $K_0$:
\beq M = C_0
(1+K_0+K_0^2+K_0^3+\ldots)\equiv \frac{C_0}{1-K_0} \; .
\eeq 
Factorization occurs by introducing the projector onto physical 
states, $\Proj$,
\beqn \frac{1}{1-K_0}
&=&\frac{1}{1-(1-\Proj )K_0-\Proj K_0} \nonumber \\ 
&\equiv& \bigg[ \frac{1}{1-(1-\Proj )K_0}\bigg]
\bigg[\frac{1}{1-\Proj K_0 \big[1-(1-\Proj )K_0 \big]^{-1}}\bigg] \; .
\eeqn 
Defining the modified kernel $K$,
\beq 
K=\frac{K_0}{1-(1-\Proj )K_0}  \; ,
\eeq
we can thus write $M$ as
\beqn 
M&=&C_0 \frac{1}{1-(1-\Proj )K_0} \frac{1}{1-\Proj K} \nonumber \\
&\equiv& C \:\: \Gamma \label{factorize}    
\eeqn 
where 
\beqn 
C&=&C_0 \frac{1}{1-(1-\Proj )K_0} \; ,\\
\Gamma &=& \frac{1}{1-\Proj K}  \; .     
\eeqn 
At this stage the factorized structure becomes apparent. In the light-cone 
gauge the 2PI
kernels $K_0$ are finite before the integration over the sides of 
the ladder is performed. Collinear singularities appear only after 
integrating over the lines connecting the rungs of the ladders[\ref{EGMPR}]. 
All collinear
singularities are contained in $\Gamma$, whereas $C$ is interpreted as the
(finite) short distance cross section. Re-expanding we find that 
\beq
K=K_0 \left( 1+(1-\Proj ) K_0 +( 1 -\Proj )(K_0 (1-\Proj )K_0) + \ldots
\right) \; ,
\eeq 
\beqn \Gamma &=&
1+\Proj K +(\Proj K)(\Proj K) + \ldots \nonumber \\ 
&\equiv & 1+\Proj K_0 +\Proj K_0 ( 1-\Proj )K_0 +(\Proj K_0)(\Proj K_0) 
+\ldots
\eeqn 
Restoring the indices and regulating collinear singularities 
by going to $d=4-2 \epsilon$ dimensions we have that
\beq \label{gamma}
\Gamma_{ij}=Z_{j} \frac{1}{1-\Proj K} = Z_{j}
\Bigg[1+\Proj K_0 +\Proj (K_0^2) - \Proj (K_0 \Proj K_0) + \ldots \Bigg] 
\eeq
and explicitly 
\beq
\Gamma_{ij}(z,\as,\frac{1}{\epsilon})=Z_{j} \Bigg[\delta(1-z) \delta_{ij}
+z \; \mbox{PP} \int \frac{d^dk}{(2 \pi)^d}
\delta(z-\frac{n \cdot k}{pn}) U_i K \frac{1}{1-\Proj K} L_j\Bigg] 
\eeq 
where `PP' extracts the pole part of the expression on its right
and $Z_j$ ($j=q(g)$) is the 
residue of the pole of the full quark (gluon) propagator, contributing to 
the diagonal splitting functions. Furthermore the spin averaged
projection operators onto physical states are given by, 
\beqn 
&&U_q =\frac{1}{4 n \cdot k}\slsh{n},\;\;\;
L_q = \slsh{p} \nonumber \\ 
&&U_g = -g^{\mu\nu},\;\;\; 
L_g = \frac{1}{d-2} \Bigg[-g^{\mu \nu}+\frac{n^\mu p^\nu+n^\nu p^\mu}{pn}
\Bigg] \; . 
\eeqn
\subsection{Derivation of GLAP equation}
\def\as{\alpha_S}
\def\half{\mbox{\small $\frac{1}{2}$}}
The property of factorization allows
us to separate the low momentum physics from the high momentum physics
in a multiplicative way. This separation is performed at a scale $\mu$,
which is completely arbitrary, and no physical prediction can depend on
it. In this section we investigate the constraints provided by this
condition. For simplicity, we will consider a non-singlet cross
section which can only be initiated by a quark.
We therefore have the factorized result,
\beq \label{factnonsing}
\sigma(\frac{Q^2}{\mu^2},\as(\mu^2),\epsilon) = 
\tilde{\sigma}_q (\frac{Q^2}{\mu^2},\as(\mu^2)) 
\otimes \Gamma_{qq} (\as(\mu^2),\epsilon)
\eeq
where we have indicated that $\Gamma_{qq}$ does {\em not} depend on $Q^2$
(i.e. $Q^2/\mu^2$), which is a consequence[\ref{CFP}] of the finiteness of 
the kernel $K_0$ in the light-cone gauge. The symbol $\otimes$ indicates a 
convolution integral over longitudinal momentum fractions of the type 
\beq
f \otimes g \equiv \int_0^1 dy \; dz \; f(y) g(z) \; \delta(x-yz)  \; .
\eeq
If we take moments,
\beq
f(j)=\int_0^1 dx x^{j-1} f(x)
\eeq
on both sides of 
Eq.~(\ref{factnonsing}), it reduces to a simple product:
\beq 
\sigma(j,\frac{Q^2}{\mu^2},\as(\mu^2),\epsilon) = 
\tilde{\sigma}_q (j,\frac{Q^2}{\mu^2},\as(\mu^2)) 
\Gamma_{qq}(j,\as(\mu^2),\epsilon)  \; .
\eeq
$\tilde{\sigma}_q$ is the short distance cross section 
from which all singularities have been factorized. $\Gamma_{qq}$ 
contains the mass
singularities which manifest themselves as poles in $\epsilon$. 
The independence of the full cross section of $\mu$ implies that
\beq
\frac{d}{d \ln \mu^2} \sigma=0
\eeq
and hence that
\beq \label{muindependence}
\frac{d}{d \ln \mu^2} \ln \Gamma_{qq} (j,\as(\mu^2),\epsilon) = 
-\frac{d}{d \ln \mu^2} \ln \tilde{\sigma}_q (j,\frac{Q^2}{\mu^2}
,\as(\mu^2)) = \gamma_{qq} (j,\as(\mu^2))   \; .
\eeq
The function $\gamma_{qq}$ is known as the anomalous dimension, because 
it measures the deviation of $\tilde{\sigma}_q$ from its naive scaling 
dimension.
It must be finite and can only depend on $\as(\mu^2)$ because 
these are the only variables common to both $\Gamma_{qq}$ 
and $\tilde{\sigma}_q$.
The anomalous dimension is extracted from Eq.~(\ref{muindependence}) 
in the following way:
Because the $\mu$ dependence of $\Gamma_{qq}$ enters only through 
the running coupling we have that,
\beq \label{muindep1}
\gamma_{qq}(j,\as(\mu^2)) \equiv \beta(\as,\epsilon)  
\frac{d}{d \as } \ln \Gamma_{qq} (j,\as(\mu^2),\epsilon)  \;  ,
\eeq
where $\beta (\as,\epsilon)$ is the $d$-dimensional QCD $\beta$ function
in the $\overline{\mbox{MS}}$ scheme,
\beq 
\beta (\as,\epsilon) = \frac{d\as}{d\ln \mu^2}=
-\epsilon \as + \beta (\as)  \; .
\eeq
In the $\overline{\mbox{MS}}$ scheme $\Gamma_{qq}$ is given by 
a series of the form
\beq
\Gamma_{qq} (j,\as,\epsilon) = 1 +\sum_{i=1}^\infty 
\frac{\Gamma_{qq}^{(i)}(j,\as)}{\epsilon^i}  \; .
\eeq
Comparing the coefficient of the term of order $\epsilon^0$ we find that
\beq
\gamma_{qq} (j,\as)= - \frac{d}{d \ln \as} \Gamma_{qq}^{(1)}(j,\as) \; .
\eeq
Integrating Eqs.~(\ref{muindependence},\ref{muindep1}) one obtains 
\beq 
\Gamma_{qq} (j,\as,\epsilon) = \exp \Bigg\{ \int_0^{\as}
d\lambda \frac{\gamma_{qq}(j,\lambda)}{\beta(\lambda)-\epsilon 
\lambda} \Bigg\}
\eeq
and 
\beq 
\sigma(j,\frac{Q^2}{\mu^2},\as(\mu^2),\epsilon) = 
\tilde{\sigma}_q (j,1,\as(Q^2)) \exp \Bigg\{ \int_0^{\as(Q^2)}
d\lambda \frac{\gamma_{qq}(j,\lambda)}{\beta(\lambda)-\epsilon 
\lambda} \Bigg\}
\eeq
with the running coupling $\as (Q^2)$. 
In order to obtain the hadronic cross section, $\sigma (Q^2/\mu^2,
\as,\epsilon)$ has to be convoluted with `bare' (`unrenormalized') 
quark densities $\tilde{q}(\as,\epsilon)$ which contain mass 
singularities that must exactly cancel those in $\Gamma_{qq}$. The
resulting `dressed' (`renormalized') quark distribution function 
\beq
q(j,Q^2) = \exp \Bigg\{ \int_0^{\as(Q^2)}
d\lambda \frac{\gamma_{qq}(j,\lambda)}{\beta(\lambda)-\epsilon 
\lambda} \Bigg\} \tilde{q}(j,\as (\mu^2),\epsilon)
\eeq
is free of mass singularities and
satisfies the non-singlet evolution equation 
\beq
\frac{d q(j,Q^2)}{d \ln Q^2}=\gamma_{qq}(j,\as (Q^2)) q(j,Q^2)
\eeq
which, defining 
\beq 
\int_0^1 dz z^{j-1} P_{qq}(z,\as) \equiv
\gamma_{qq} (j,\as)   \; ,
\eeq
in $x$ space takes the form of the GLAP equation[\ref{AlPa},\ref{GLAP}]: 
\beq \label{mastereq}
\frac{d q(x,Q^2) }{d \ln Q^2 } 
=\int_0^1 dy \int_0^1 dz \; P_{qq}(y,\as (Q^2)) q(z,Q^2) \; \delta (x-yz) 
\; .
\eeq
Expanding
\beq \label{expan}
P_{qq} (x,\as) = \left( \frac{\as}{2\pi} \right) P_{qq}^{(0)} (x) + 
\left( \frac{\as}{2\pi} \right)^2 P_{qq}^{(1)} (x) + \ldots
\eeq
one has
\beq
\Gamma_{qq} (x,\as,\epsilon) = \delta (1-x) - \frac{1}{\epsilon}
\Bigg(\frac{\as}{2\pi} P_{qq}^{(0)}(x)+\frac{1}{2} \left( \frac{\as}{2\pi}
\right)^2 P_{qq}^{(1)} (x) + \ldots \Bigg) + O \left(
\frac{1}{\epsilon^2} \right) \; .
\eeq
The generalization to the singlet case is straightforward.
\subsection{Non-singlet and singlet equations}
The separation 
into singlet and non-singlet parts depends on the properties of the 
kernel.  Using $SU(f)$ flavour symmetry we may define the following 
combinations of $qq$ and $q \bar{q}$ matrix elements:
\beqn \label{symm}
P_{q_iq_k}&=&\delta_{ik} P^V_{qq}+P^S_{qq} \nonumber \\
P_{q_i\bar{q}_k}&=&\delta_{ik} P^V_{q\bar{q}}+P^S_{q\bar{q}}\nonumber \\
P^{\pm}&=&P^V_{qq} \pm P^V_{q\bar{q}}   \; .
\eeqn
In addition, because of 
charge conjugation invariance, we have that
\beqn \label{kern}
P_{q_iq_j}&=&P_{\bar{q}_i\bar{q}_j} \nonumber \\
P_{q_i \bar{q}_j}&=&P_{\bar{q}_i q_j} \nonumber \\
P_{q_i g}&=&P_{\bar{q}_i g} \nonumber \\
P_{g q_i}&=&P_{g \bar{q}_i} \; .
\eeqn

At two loop order, there is a non-zero contribution from
$P^S_{qq}$ and $P^S_{q\bar{q}}$, but we have the additional relation
\beq \label{simp}
P^S_{qq}=P^S_{q\bar{q}} \; .
\eeq
which simplifies the treatment of the non-singlet pieces.

For each flavour we define the sum and difference of the 
quark and anti-quark distributions as 
\beq \label{qplus}
 q^\pm_i = q_i \pm \bar{q}_i   \; .
\eeq
One then finds that the combinations 
\beq 
V_i = q_i^-
\eeq
and 
\beq
T_l = \sum_{i=1}^k q_i^+ - k q_k^+
\eeq
(where $i,k=1,\ldots ,n_f$; $\; l=k^2-1$) are non-singlets, i.e., 
evolve according to 
Eq.~(\ref{mastereq}) with the kernels $P^-$ and $P^+$, respectively. 

The singlet Altarelli-Parisi equation is\footnote{Note that the notation 
for the off-diagonal terms is different than in 
ref.~[\ref{FP}].}[\ref{AlPa},\ref{GLAP}] 
\begin{eqnarray} \label{AP}
\frac{d }{d \ln Q^2}   
\left( \begin{array}{c}\Sigma (j,Q^2) \\ G(j,Q^2) \end{array} \right)
= \left( \begin{array}{cc} P_{qq}(j,\as (Q^2)) &  P_{qg}(j,\as (Q^2))\\  
P_{gq}(j,\as (Q^2) ) &  P_{gg}(j,\as (Q^2) )\\  
\end{array}\right)
\left(  \begin{array}{c}
\Sigma(j,Q^2) \\ 
 G(j,Q^2) 
\end{array} \right)
\end{eqnarray}
where $G(j)$ is the moment of the gluon distribution and $\Sigma(j)$ is 
the singlet quark combination,
\beq
\Sigma(j,Q^2) = \sum_f  q_i^+(j,Q^2) \equiv 
 \sum_f  \left[ q_i(j,Q^2) + \bar{q}_i(j,Q^2) \right]   \; .
\eeq
The elements of the anomalous dimension matrix are 
given in terms of the kernels defined in Eqs.~(\ref{symm}-\ref{simp}) as,
\beqn
 P_{qq}&=&P^+ + n_f (P^S_{qq} +P^S_{q\bar{q}})
 \nonumber \\
 P_{qg}&=&2 n_f P_{q_ig}    \nonumber \\
 P_{gq}&=&P_{gq_i} \; .
\eeqn
\subsection{Renormalization constants}
\begin{figure}[htb]
\def\capa{Renormalization constants and the vertices which they renormalize}
\vspace{8cm}
\includegraphics{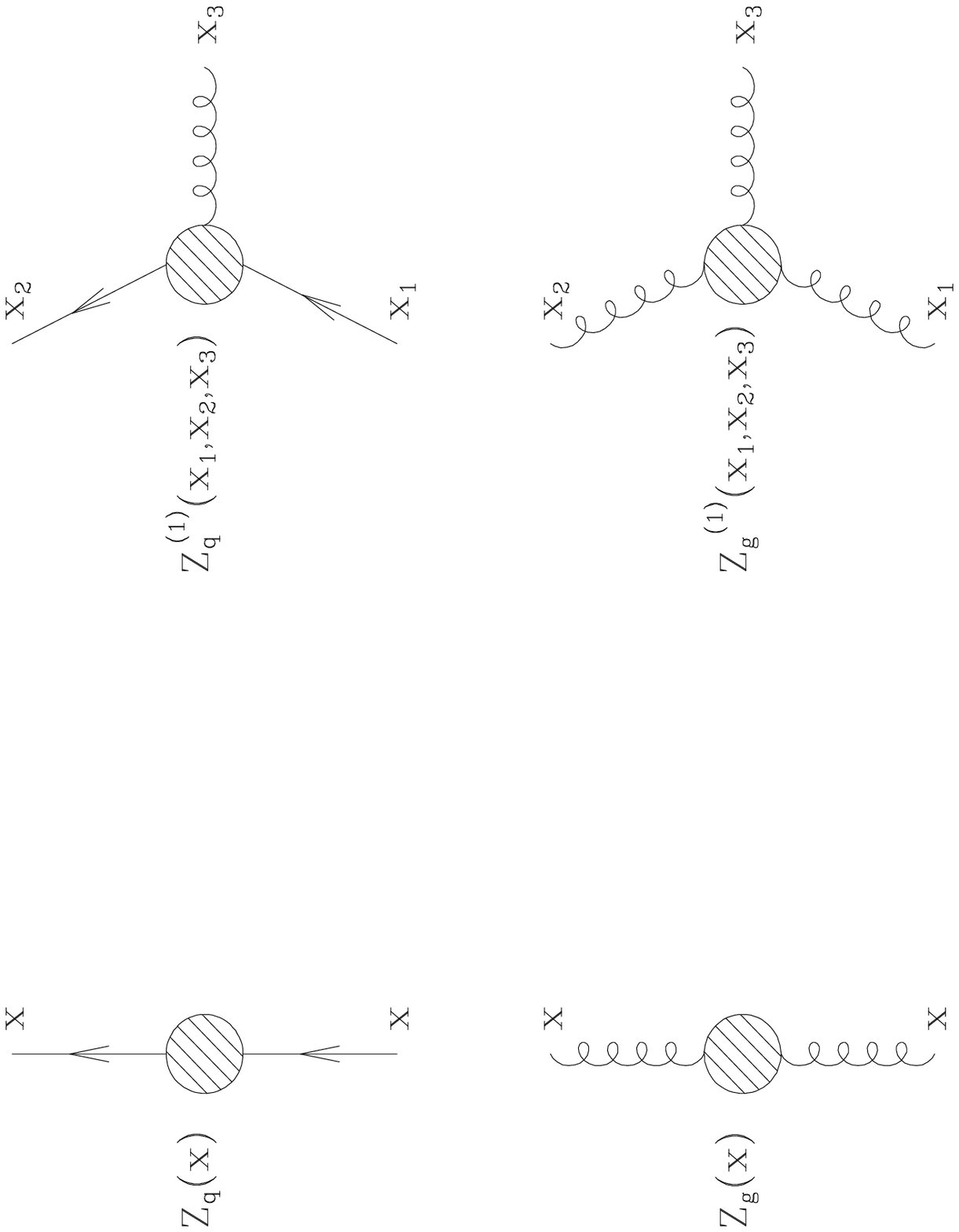}
\caption{\capa}
\label{renorm}
\end{figure}

The notation for the renormalization constants is shown in 
Fig.~\ref{renorm}. We define the integral 
\beq \label{i0def}
I_0 = \int_0^1 du \frac{u}{u^2+\delta^2} 
\eeq
which contains the divergences in the PV regulator $\delta$
(see Eq.~(\ref{PPprescription})) arising from the light-cone gauge 
propagator. As already noted in ref.~[\ref{CFP}], use of the 
PV prescription~(\ref{PPprescription}) in the light-cone gauge entails
the disagreeable feature that the renormalization constants 
depend on the longitudinal momentum fractions $x$. 
\beqn \label{rencon} 
Z_q(x)
 &=& 1+\frac{\as}{2 \pi} \frac{1}{2\epsilon} 
  \Bigg[C_F (-3+4 I_0+4 \ln x)  \Bigg] \nonumber \\
Z_g(x) &= & 1+\frac{\as}{2 \pi} \frac{1}{2\epsilon}
 \Bigg[\frac{4 T_f}{3}+\nc (-\frac{11}{3}+4 I_0+4 \ln x) \Bigg] \nonumber \\
Z_q^{(1)}(x_1,x_2,x_3)
 &=& 1+\frac{\as}{2 \pi} \frac{1}{2\epsilon}
  \Bigg[C_F (3-4 I_0-2 \ln x_1-2 \ln x_2) -2 \nc (I_0+\ln x_3)  \Bigg] 
\nonumber \\
Z_g^{(1)}(x_1,x_2,x_3)
 &=& 1+\frac{\as}{2 \pi} \frac{1}{2\epsilon}
  \Bigg[\nc (\frac{11}{3}-6 I_0-2 \ln x_1-2\ln x_2-2\ln x_3)
 -\frac{4 T_f}{3} \Bigg] \nonumber \\
\eeqn
where 
\beq
C_F=\frac{4}{3},\;
\nc=3,\;
T_f=T_R n_f = \half n_f \; .
\eeq
When combined in the appropriate combinations to investigate the 
renormalization of the charge, the momentum dependent terms and the 
divergent integrals $I_0$ cancel. Thus the relationship 
between the bare and renormalized couplings is 
\beqn
\as^{(0)}&=&\as\mu^{2\epsilon} Z_q^{(1)}(x_1,x_2,x_3) 
\sqrt{Z_q(x_1)Z_q(x_2)Z_g(x_3)} \nonumber \\
         &=&\as\mu^{2\epsilon} Z_g^{(1)}(x_1,x_2,x_3) 
\sqrt{Z_g(x_1)Z_g(x_2)Z_g(x_3)} \\
         &=& \as \mu^{2\epsilon}
\Big[1- \frac{\as}{2 \pi} \frac{1}{2 \epsilon} 
(\frac{11 \nc}{6}-\frac{2 T_f}{3}) + \ldots \Big]
\equiv \as \mu^{2\epsilon}
\Big[1 - \frac{\as}{2 \pi} \frac{1}{4 \epsilon} \beta_0 + \ldots \Big] \; .
\nonumber 
\eeqn
\subsection{Topologies of NLO graphs}
\begin{figure}[htb]
\vspace{7cm}
\includegraphics{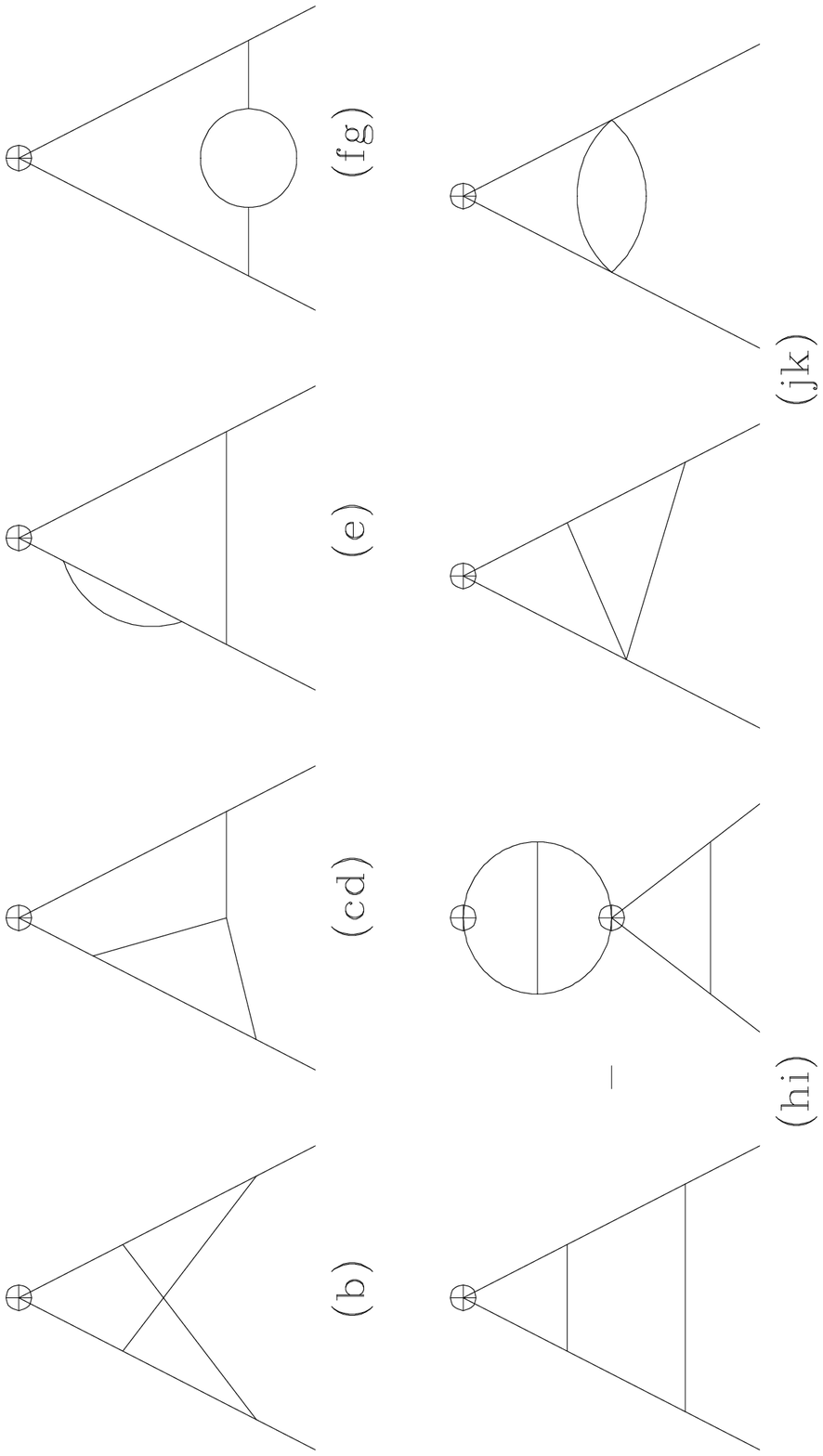}
\def\capb{Basic topologies of the diagrams}
\caption{\capb}
\label{topol}
\end{figure}

\begin{figure}[htb]
\vspace{7cm}
\includegraphics{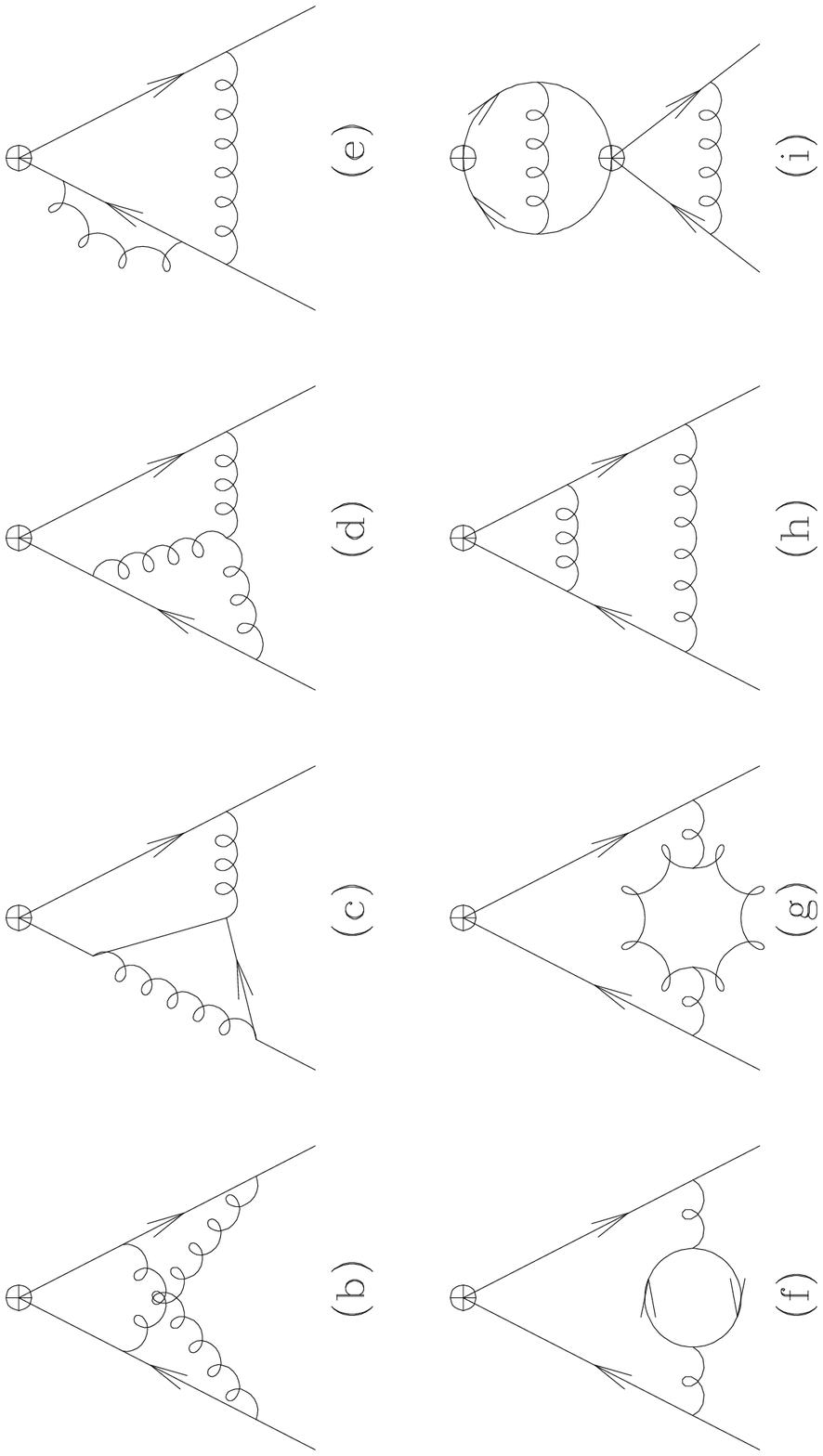}
\def\capc{Diagrams for the $qq$ part of the NLO non-singlet 
splitting functions}
\caption{\capc}
\label{qq}
\end{figure}

\begin{figure}[htb]
\vspace{4cm}
\includegraphics{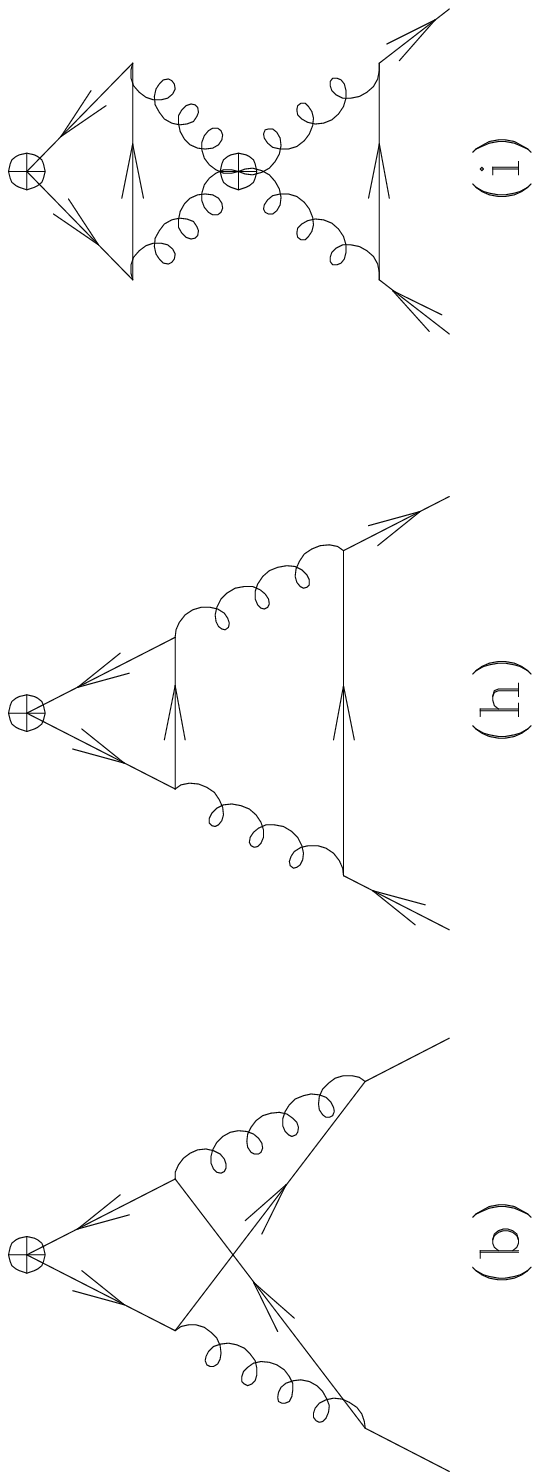}
\def\capd{The $q\bar{q}$ diagrams}
\caption{\capd}
\label{qqb}
\end{figure}

The basic topologies of all 2PI diagrams which occur 
in two loops are shown in Fig.~\ref{topol}. The notation of the topologies 
(b)-(i) is determined by the labelling of the diagrams for the non-singlet 
calculation given in ref.~[\ref{CFP}]. We have not included diagrams which 
can be obtained by reflection about the vertical axis which 
occur in cases (c),(d),(e) and (j). Topologies (hi) correspond
to the terms $\Proj (K_0^2)-\Proj (K_0 \Proj K_0)$ in Eq.~(\ref{gamma}),
all other topologies belong to $\Proj K_0$.
As an example, the diagrams corresponding to $P_{qq}^V$ are given 
explicitly in Fig.~\ref{qq}. Fig.~\ref{qqb} shows the diagrams for 
$P_{q\bar{q}}^V$ (b) and for $P_{q\bar{q}}^S$ (h,i). 
The appendices give the necessary ingredients needed for the evaluation 
of the real and virtual graphs in Fig.~\ref{topol}. 
\section{Results}
As in Eq.~(\ref{expan}) we define the perturbative expansion 
\beq
P_{ij}(x,\as)=\Big( \frac{\as}{2 \pi}\Big) P_{ij}^{(0)}(x)
 +\Big(\frac{\as}{2 \pi}\Big)^2 P_{ij}^{(1)}(x) + \ldots 
\eeq
The full one loop results are included for completeness[\ref{AlPa}]
\beqn
\label{Pqq}
P_{qq}^{(0)}(x)&=&C_F \Big\{\frac{2}{[1-x]_+}-1-x +\frac{3}{2}\delta(1-x)
\Big\} \\
\label{Pqg}
P_{qg}^{(0)}(x)&=&2 T_f\Big\{x^2+(1-x)^2\Big\}  \\
\label{Pgq}
P_{gq}^{(0)}(x)&=&C_F \Big\{\frac{1+(1-x)^2}{x} \Big\} \\
\label{Pgg}
P_{gg}^{(0)}(x)&=&2 \nc 
\Big\{\frac{1}{[1-x]_+}+\frac{1}{x} -2 +x(1-x)  \Big\}
+\frac{\beta_0}{2} \delta(1-x)  \; .
\eeqn
In order to write the full result for the two loop splitting functions
$P_{ij}^{(1)}$ we introduce the notation
\beqn
\label{pqq}
p_{qq}(x)&=&\frac{2}{1-x}-1-x \\
\label{pqg}
p_{qg}(x)&=&x^2+(1-x)^2  \\
\label{pgq}
p_{gq}(x)&=&\frac{1+(1-x)^2}{x}  \\
\label{pgg}
p_{gg}(x)&=& \frac{1}{1-x}+\frac{1}{x} -2 +x(1-x)  \; .
\eeqn
The graph-by-graph results are given in Tables~\ref{table_qqs}-4,
where we only list the contributions to the singlet splitting
functions since those for the non-singlet case have been presented in
a similar table in ref.~[\ref{CFP}].
\renewcommand\arraystretch{1.2}
\begin{table} 
\begin{center}
\begin{tabular}{|l|c|} 
\hline
&\multicolumn{1}{|c|}{$C_F T_R$} \\
\hline
      Terms                         &(hi) \\
\hline 
 $x \;  \ln^2(x)$                         &-1        \\
 $ \ln^2(x)$                              &-1        \\
 $x^2 \; \ln(x)$                          &8/3         \\
 $x \ln(x)$                               &5     \\
 $ \ln(x)$                                &1   \\
 $x^2$                                    &-56/9         \\
 $x$                                      &6     \\
 $1$                                      &-2     \\
 $1/x$                                    &20/9         \\
\hline
\hline
\end{tabular}
\caption{Results for the $q\bar{q}$-Singlet diagram}
\label{table_qqs}
\end{center}
\end{table}
\renewcommand\arraystretch{1.}
Our final full results for the two loop non-singlet splitting functions 
read for $x \neq 1$ [\ref{CFP}]:
\beqn \label{pqq2}
&&P_{qq}^{V,(1)}=
  C_F^2 \Big\{ -\big[2 \ln x \ln(1-x)+\frac{3}{2} \ln x  \big] p_{qq}(x) 
\nonumber \\ &&
 -(\frac{3}{2}+\frac{7}{2} x)\ln x
      -\frac{1}{2} (1+x) \ln^2 x -5 (1-x)\Big\}
\nonumber \\ &&
      +C_F \nc \Big\{ \big[\frac{1}{2} \ln^2 x
 +\frac{11}{6} \ln x+\frac{67}{18}-\frac{\pi^2}{6} \big] p_{qq}(x)
      +(1+x) \ln x+\frac{20}{3} (1-x)\Big\}
\nonumber \\ &&
      +
C_F T_f \Big\{-\big[\frac{2}{3} \ln x+\frac{10}{9}\big] p_{qq}(x) 
 -\frac{4}{3}  (1-x))\Big\} \\
&&P_{q\bar{q}}^{V,(1)} = C_F (C_F-\frac{\nc}{2}) 
\Big\{2 p_{qq}(-x) S_2(x)+2 (1+x) \ln x+4 (1-x) \Big\}   \; .
\eeqn
Our results for the singlet terms are[\ref{FP}],
\beqn \label{pff2}
&&P_{qq}^{(1)}=P^{+,(1)}   \\
&& + 2 C_F T_f \Big\{ \frac{20}{9x} -2 +6 x-\frac{56}{9} x^2
+(1+5 x+\frac{8}{3} x^2) \ln x - (1+x) \ln^2 x \Big\} \nonumber
\eeqn
\beqn \label{pqg2}
&&P_{qg}^{(1)}=C_F T_f
       \Big\{4-9 x-(1-4 x) \ln x-(1-2 x) \ln^2x +4 \ln(1-x)        
\nonumber \\ &&
      +\big[2 \ln^2\big(\frac{1-x}{x}\big)-4 \ln\big(\frac{1-x}{x}\big)
      -\frac{2}{3} \pi^2 +10 \big] p_{qg}(x)\Big\} \nonumber \\ &&      
      +\nc T_f
       \Big\{\frac{182}{9}+\frac{14}{9} x+\frac{40}{9 x}
  +(\frac{136}{3} x-\frac{38}{3}) \ln x-4 \ln(1-x) 
      -(2+8 x) \ln^2x \nonumber \\ && 
       +\big[-\ln^2x+\frac{44}{3} \ln x-2 \ln^2(1-x)+4 \ln(1-x)
+\frac{\pi^2}{3}
 -\frac{218}{9}\big] p_{qg}(x)
 \nonumber \\ &&
  +2 p_{qg}(-x) S_2(x) \Big\}
\eeqn
\beqn \label{pgq2}
&&P_{gq}^{(1)}=C_F^2 \Big\{-\frac{5}{2}-\frac{7}{2} x +(2+\frac{7}{2} x) 
 \ln x-(1-\frac{1}{2} x)
       \ln^2x-2 x \ln(1-x)
 \nonumber \\ &&
 -\big[ 3 \ln(1-x)+\ln^2(1-x)\big] p_{gq}(x)\Big\}
 \nonumber \\ &&
      +C_F \nc \Big\{
  \frac{28}{9}+\frac{65}{18} x+\frac{44}{9} x^2
 -(12+5 x+\frac{8}{3} x^2) \ln x    
      +(4+x) \ln^2x+2 x \ln(1-x)   \nonumber \\ &&
      +\big[ -2 \ln x \ln(1-x)+\frac{1}{2} \ln^2x+\frac{11}{3}      
       \ln(1-x)+\ln^2(1-x)-\frac{\pi^2}{6}+\frac{1}{2}\big] p_{gq}(x)
 \nonumber \\ &&
  +S_2(x) p_{gq}(-x)\Big\} \nonumber \\ &&
      +C_F T_f \Big\{
    -\frac{4}{3} x-\big[\frac{20}{9}+\frac{4}{3} \ln(1-x) \big] 
p_{gq}(x)\Big\}
\eeqn
\beqn \label{pgg2}
&&P_{gg}^{(1)}=C_F T_f \Big\{-16+8 x+\frac{20}{3}x^2 +\frac{4}{3 x}
      -(6+10 x) \ln x- 2 (1+x) \ln^2 x \Big\}
  \nonumber \\ &&
      +\nc T_f
       \Big\{ 2-2 x+\frac{26}{9} (x^2-\frac{1}{x})
 -\frac{4}{3} (1+x) \ln x-\frac{20}{9} p_{gg}(x) \Big\}\nonumber \\ && 
      +\nc^2 \Big\{ \frac{27}{2} (1-x)+\frac{67}{9} (x^2-\frac{1}{x}) 
      -(\frac{25}{3}-\frac{11}{3} x+\frac{44}{3} x^2) \ln x 
      +4 (1+x) \ln^2 x \nonumber \\ && 
  + \big[ \frac{67}{9}-4 \ln x \ln(1-x)       
      +\ln^2 x  -\frac{\pi^2}{3}\big] p_{gg}(x)+2 p_{gg}(-x) S_2(x)) \Big\}
\eeqn
where the function $S_2(x)$ is defined as\footnote{Note that the definition 
of $S_2$ in ref.~[\ref{FP}] contains a typographical mistake.}
\beq
S_2(x)= \int_{\frac{x}{1+x}}^{\frac{1}{1+x}} \frac{dz}{z} 
\ln \big(\frac{1-z}{z}\big)  \; .
\eeq
In the small-$x$ limit $S_2$ becomes
\beq
S_2=\frac{1}{2} \ln^2 x - \frac{\pi^2}{6}+O(x)  \; .
\eeq
All results in Eqs.~(\ref{pqq2}-\ref{pgg2}) are in complete agreement
with the corresponding results in [\ref{CFP},\ref{FP}].
They can be extended to all values of $x$ using a trick to 
evaluate the endpoint contributions in Eqs.~(\ref{pqq2},\ref{pgg2}). 
The sum rule from the conservation of fermion number is
\beq
\int_0^1 dx \; \Big( P_{qq}(x)- P_{q\bar{q}}(x) \Big) \equiv 
\int_0^1 dx \;  P^-(x) =0 \; .  
\eeq
The conservation of momentum leads to the following two relations,
\beqn
\int_0^1 dx \; x \Big( P_{qq}(x)+P_{gq}(x) \Big) =0  \; , \\
\int_0^1 dx \; x \Big( P_{qg}(x)+P_{gg}(x) \Big) =0  \; .
\eeqn
These results for the integrals of the splitting functions 
are satisfied if one makes the substitutions
\beq \label{plusify}
\frac{1}{1-x} \rightarrow \frac{1}{[1-x]_+} 
\eeq
in Eqs.~(\ref{pqq},\ref{pgg}) and adds in the end-point contributions to 
Eqs.~(\ref{pqq2},\ref{pgg2}) [\ref{HeWa}],
\beqn
P_{qq}^{V,(1)} \rightarrow P_{qq}^{V,(1)}
&+&\Bigg[ 
C_F^2 \Big\{\frac{3}{8}-\frac{\pi^2}{2}+6 \zeta(3) \Big\}
+C_F \nc \Big\{\frac{17}{24}+\frac{11\pi^2}{18}-3 \zeta(3)\Big\}
\nonumber \\ && 
-C_F T_f \Big\{\frac{1}{6}+\frac{2 \pi^2}{9}\Big\}\Bigg]\delta(1-x) \\
P_{gg}^{(1)} \rightarrow P_{gg}^{(1)}
&+&\Bigg[\nc^2 \Big\{\frac{8}{3}+3 \zeta(3)\Big\}
-C_F T_f -\frac{4}{3} \nc T_f \Bigg]\delta(1-x)
\eeqn
where $\zeta(3) \approx 1.202057$. The substitution in Eq.~(\ref{plusify}) 
is obviously not necessary if the factor of $1/(1-x)$ has a coefficient 
which vanishes at $x=1$, such as $\ln(x)$.
\begin{figure}[htb]
\vspace{16cm}
\includegraphics{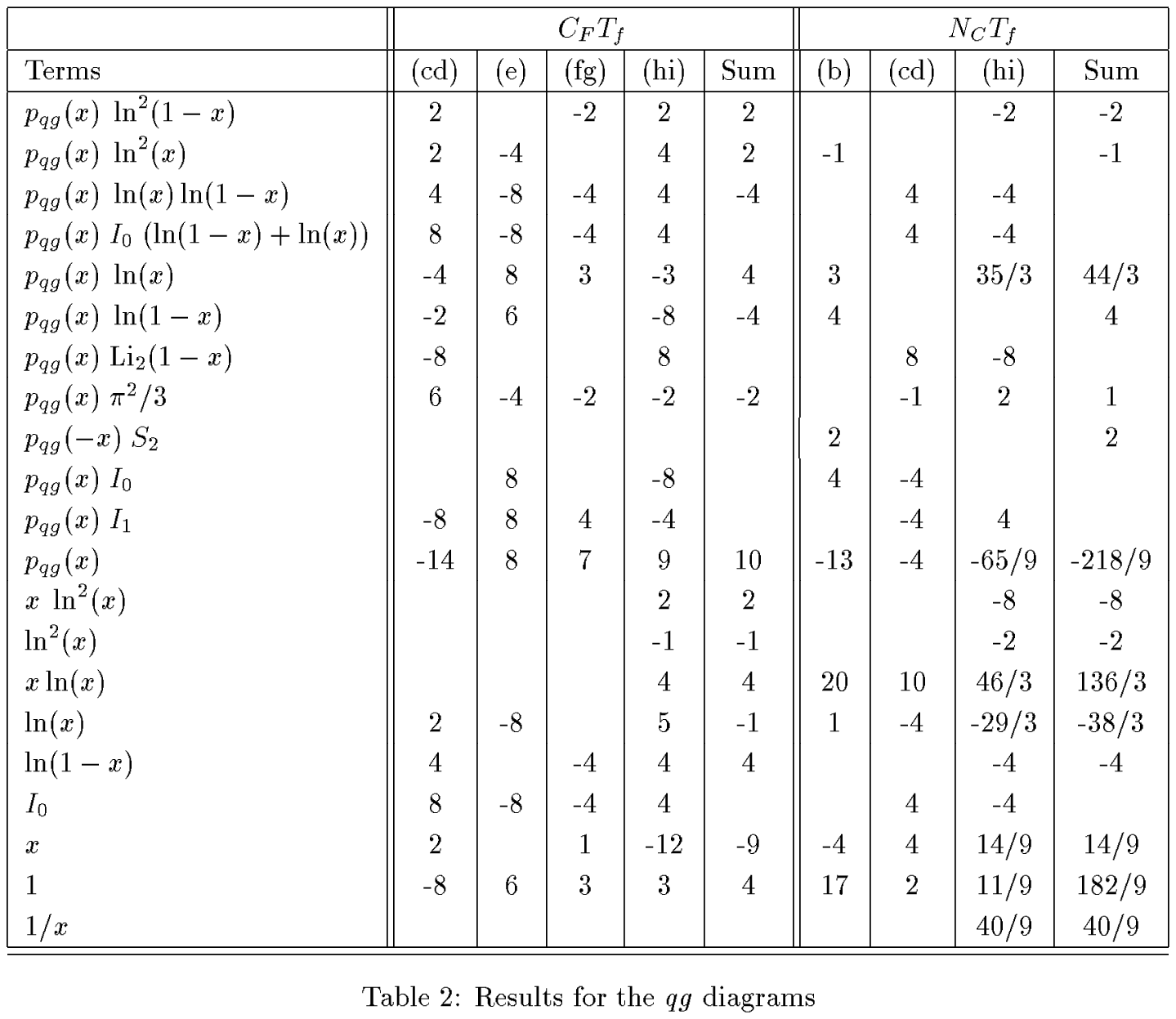}
\end{figure}
\begin{figure}[htb]
\vspace{16cm}
\includegraphics{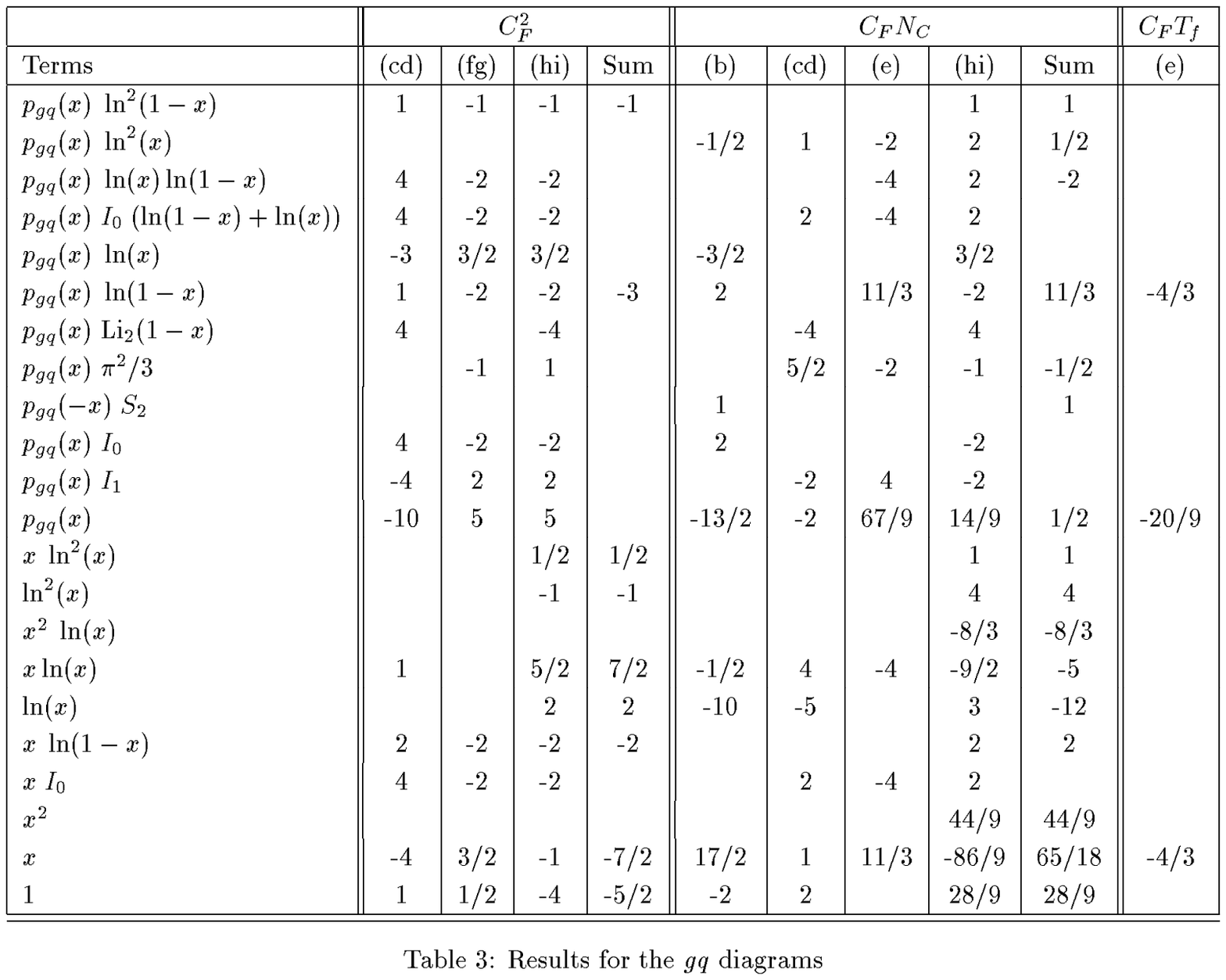}
\end{figure}
\begin{figure}[htb]
\vspace{16cm}
\includegraphics{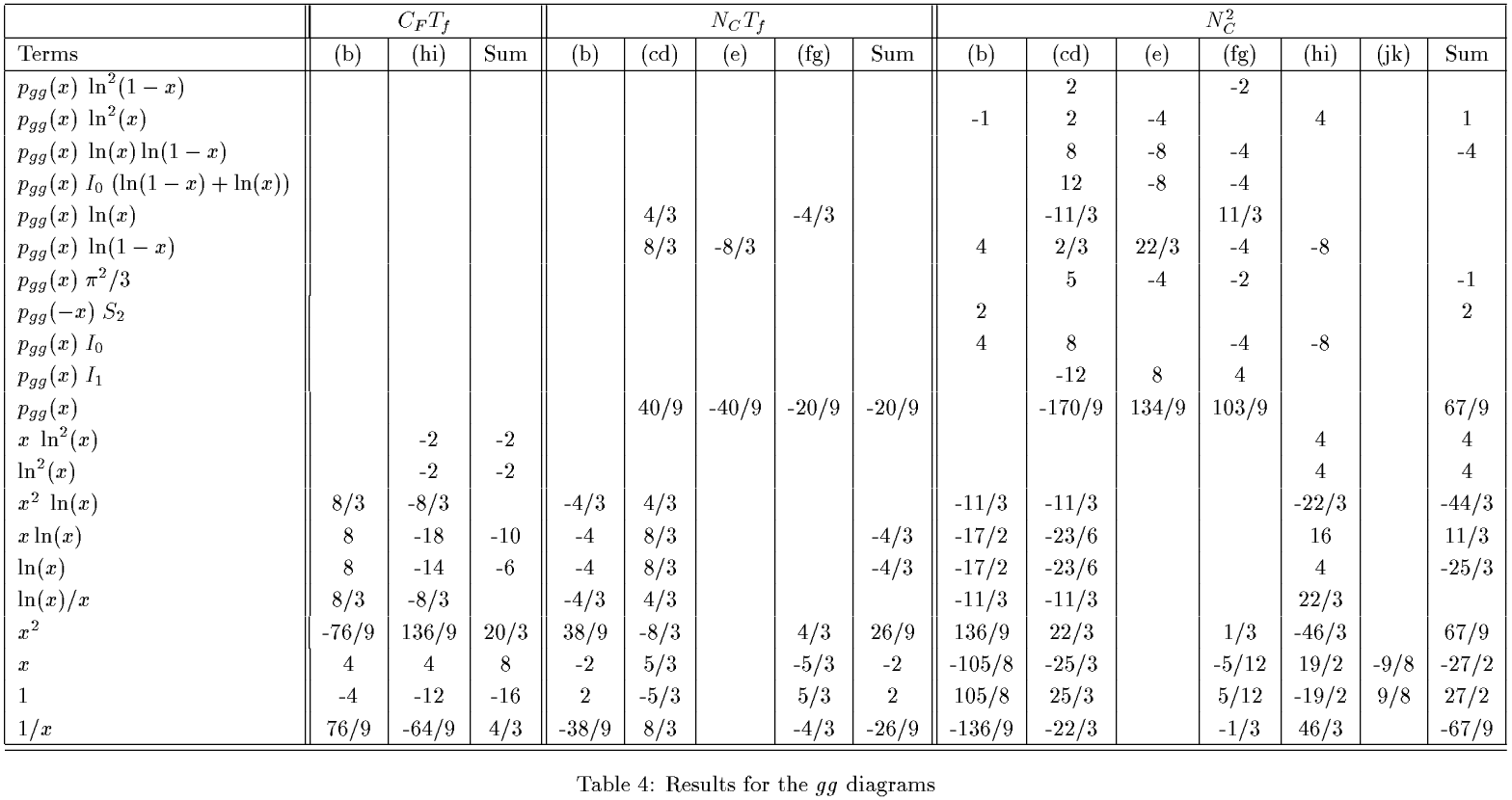}
\end{figure}
\setcounter{table}{6}
\clearpage
\section{Summary}
This paper has presented a recalculation of the two loop anomalous dimension  
for space-like processes. The results given in this paper are therefore 
not new. The new features are the presentation of the results in a 
coherent notation, the description of some of the integrals which are 
required to derive the results and a detailed description of the contributions 
of the sub-diagrams to the results. 
\appendix
\section{Virtual Integrals}
\setcounter{equation}{0}
\renewcommand{\theequation}{\thesection.\arabic{equation}}
\subsection{Two point function}
The evaluation of the virtual integrals involving non-covariant
denominators of the form $1/(n \cdot k)$ requires some care. We define
\beqn                                
\lp & =& n\cdot l \nonumber \\
\lm & =& p\cdot l \nonumber \\
d^d l &=& d\lp d\lm d^{d-2} \l_\perp  \; .
\eeqn
We shall evaluate the integrals by explicitly performing the integrals 
over $\lm,l_\perp$ keeping $\lp$ fixed. This formulation will be useful 
when $f(\lp)$ contains poles in $1/\lp$ coming from non-covariant denominators
as long as the method used to regulate the $\lp$ singularity does not 
involve $\lm$. The general two-point function then reads
\beqn
&&\int \frac{d^d l}{(2 \pi)^d} \frac{f(\lp)}
{(l^2+i \varepsilon)((l-k)^2+i \varepsilon)} 
\nonumber \\
&&= \frac{i}{16 \pi^2}  \left(\frac{4 \pi}{-k^2} \right)^\epsilon
\frac{\Gamma(1+\epsilon)}{\epsilon} 
 \int_{0}^{1} dz \; f(\lp) \; z^{-\epsilon} (1-z)^{-\epsilon}
\eeqn
where $f$ is an arbitrary function, $d=4-2 \epsilon$ and 
$z= \lp/\kp$ the boost invariant rescaled value of $\lp$.
If $f(\lp)=1$ we recover the normal covariant result,
\beq
J_2\equiv \int \frac{d^d l}{(2 \pi)^d} \frac{1}
{(l^2+i \varepsilon)((l-k)^2+i \varepsilon)} 
 = \frac{i}{16 \pi^2} 
\left(\frac{4 \pi}{-k^2} \right)^\epsilon
\frac{\Gamma(1+\epsilon)\Gamma^2(1-\epsilon) }{\Gamma(2-2 \epsilon)} 
\frac{1}{\epsilon}  \; .
\eeq
The result for the integral with one non-covariant denominator is
\begin{eqnarray}
J_{2,n} &\equiv&
\int \frac{d^d l}{(2 \pi)^d} \mbox{PV}\Bigg( \frac{\kp}{\lp-\kp }\Bigg)
 \frac{1} {(l^2+i \varepsilon)((l-k)^2+i \varepsilon)} \nonumber \\
&&= \frac{i}{16 \pi^2} 
\left(\frac{4 \pi}{-k^2} \right)^\epsilon
\frac{\Gamma(1+\epsilon) \Gamma^2(1-\epsilon) }{\Gamma(1-2 \epsilon)}
 \Bigg\{ -\frac{I_0  + \ln x  }{\epsilon} \nonumber \\
&&
 -I_0 \ln x +I_1 -\frac{1}{2} \ln^2 x -\frac{\pi^2}{6} +O(\epsilon) \Bigg\} 
\end{eqnarray}
where $x=n\cdot k/pn$ and we have indicated that the PV prescription 
defined by Eq.~(\ref{PPprescription}) has been used.
$I_0$ is defined as in Eq.~(\ref{i0def}); in the small-$\delta$
limit it reduces to
\beq
I_0 = \int_0^1 du \frac{u}{u^2+\delta^2} \approx -\ln |\delta| \; .
\eeq
Furthermore,
\beq
I_1 = \int_0^1 du \frac{u \ln u}{u^2+\delta^2} \approx 
-\frac{1}{2} \ln^2 |\delta| -\frac{\pi^2}{24}  \; .
\eeq

For the case without an endpoint singularity in the integral over the 
plus component we may take the limit $\delta \rightarrow 0$ and hence obtain,
\begin{eqnarray}
&&\int \frac{d^d l}{(2 \pi)^d} \frac{\kp}{\lp-\pp }
 \frac{1} {(l^2+i \varepsilon)((l-k)^2+i \varepsilon)} \nonumber \\
&&= \frac{i}{16 \pi^2} 
\left(\frac{4 \pi}{-k^2} \right)^\epsilon
\frac{\Gamma(1+\epsilon) \Gamma^2(1-\epsilon) }{\Gamma(1-2 \epsilon)}
 \Bigg\{ \frac{\ln (1-x)  }{\epsilon}
 +2 \; {\rm Li}_{2}(1-x)-\frac{\pi^2}{3}\nonumber \\
&& +2 \ln x \ln (1-x)-\frac{1}{2} \ln^2(1-x) 
+O(\epsilon) \Bigg\} 
\end{eqnarray}
where ${\rm Li}_{2}(x)$ is the usual dilogarithm function,
\beq
{\rm Li}_{2}(x) = -\int_0^x \frac{\ln(1-t)}{t} dt  \; .
\eeq
We note that one also needs the vector two point function with
one non-covariant denominator, 
\beq
J_{2,n}^{\mu} \equiv 
\int \frac{d^d l}{(2 \pi)^d} \mbox{PV}\Bigg( \frac{\kp}{\lp}\Bigg)
\frac{l^{\mu}} {(l^2+i \varepsilon)((l-k)^2+i \varepsilon)}  \; .
\eeq
Assuming Lorentz covariance of the momentum integral one finds
\beq
J_{2,n}^{\mu} = \left( k^{\mu} - \frac{k^2}{n \cdot k} n^{\mu} \right) J_2 + 
\frac{k^2}{2 n \cdot k} n^{\mu} \left( - J_{2,n} + \frac{1}{k^2} 
J_{1,n} \right)
\eeq
where 
\beq
J_{1,n} = 
\int \frac{d^d l}{(2 \pi)^d} \mbox{PV}\Bigg( \frac{\kp}{\lp}\Bigg)
\frac{1} {((l-k)^2+i \varepsilon)}   \; .
\eeq
It turns out that the integral $J_{1,n}$ always cancels out in the 
final answer.
\begin{figure}[htb]
\vspace{5cm}
\includegraphics{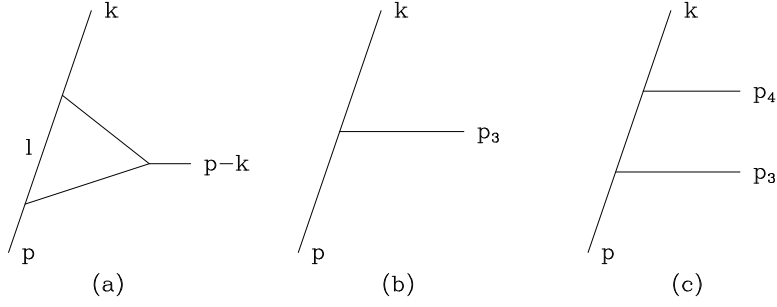}
\caption{(a) Vertex correction graph (b) One parton emission (c) 
Two parton emission.}
\label{pd}
\end{figure}
\subsection{Three point function}
We shall only consider the special case which is needed for our purpose.
We employ the momentum assignments $p^2=(p-k)^2=0$ and define the
boost invariant quantities, $x=\kp/\pp,y=\lp/\pp, z=y/x=\lp/\kp$. 
The corresponding diagram is shown in Fig.~\ref{pd}a. One finds
\begin{eqnarray} \label{basicthreepoint}
&&\int \frac{d^d l}{(2 \pi)^d} \frac{f(\lp)}
{(l^2+i \varepsilon)((l-k)^2+i \varepsilon)
((l-p)^2+i \varepsilon)} 
 = \frac{-i}{16 \pi^2 k^2} 
\left(\frac{4 \pi}{-k^2} \right)^\epsilon
\frac{\Gamma(1+\epsilon)}{\epsilon} \nonumber \\
&& \Bigg[ \int_{0}^{x} d y \; f(\lp) \; 
z^{-\epsilon} (1-z)^{-1-\epsilon} \;
{}_2 F_1 
\left( 1+\epsilon,1;1-\epsilon;\frac{z(1-x)}{z-1} \right)\nonumber \\
&&+ 2 \frac{\Gamma^2 (1-\epsilon)}{\Gamma (1-2 \epsilon)} 
(1-x)^{\epsilon} \int_{x}^{1} d y \; f(\lp) \; 
(1-y)^{-1-2\epsilon} \Bigg]
\end{eqnarray}
where ${}_2 F_1$ is the hypergeometric function. 
With a little work one can show that 
for the special case $f(\lp)=1$ one recovers the normal covariant result,
\begin{eqnarray}
&&\int \frac{d^d l}{(2 \pi)^d} \frac{1}
{(l^2+i \varepsilon)((l-k)^2+i \varepsilon)
((l-p)^2+i \varepsilon)} \nonumber \\
&& = \frac{i}{16 \pi^2 k^2} 
 \left(\frac{4 \pi}{-k^2} \right)^\epsilon
\frac{\Gamma(1+\epsilon)}{\epsilon^2} 
\frac{\Gamma^2(1-\epsilon) }{\Gamma(1-2 \epsilon)}  \; .
\end{eqnarray}
The explicit result for the three point function with one light-cone gauge 
denominator using the PV prescription is 
(we have performed a shift of the integration variables relative 
to Eq.~(\ref{basicthreepoint})):
\begin{eqnarray} \label{3dennoncovint}
&&\int \frac{d^d l}{(2 \pi)^d} \; \mbox{PV}\left(\frac{\pp}{\lp+\pp}\right)
\frac{1}{((l+p)^2+i \varepsilon)(l^2+i \varepsilon)
((l+p^\prime)^2+i \varepsilon)}= 
\nonumber \\
&& \frac{i}{16 \pi^2 k^2} 
\left(\frac{4 \pi}{-k^2} \right)^\epsilon 
\frac{\Gamma(1+\epsilon)\Gamma^2(1-\epsilon) }{\Gamma(1-2 \epsilon)}
 \Bigg\{\frac{1}{\epsilon^2}+\frac{ \ln x -I_0}{\epsilon}
\nonumber \\
&& +I_1-I_0 \ln x  -2 {\rm Li}_{2}(1-x) -\frac{1}{2} \ln^2 x - \frac{\pi^2}{6}
+O(\epsilon) \Bigg\}  \; .
\end{eqnarray}
The other integral which we need can be obtained from 
Eq.~(\ref{3dennoncovint}) by exchange of $p$ and $p^\prime$
\begin{eqnarray}
&&\int \frac{d^d l}{(2 \pi)^d} \; \mbox{PV}\left(\frac{\ppp}{\lp+\ppp} \right)
\frac{1}{((l+p)^2+i \varepsilon)(l^2+i \varepsilon)
((l+p^\prime)^2+i \varepsilon)}= 
\nonumber \\
&& \hspace*{-0.5cm} \frac{i}{16 \pi^2 k^2} 
\left(\frac{4 \pi}{-k^2} \right)^\epsilon 
\frac{\Gamma(1+\epsilon)\Gamma^2(1-\epsilon) }{\Gamma(1-2 \epsilon)}
 \Bigg\{ \frac{1}{\epsilon^2}+\frac{ \ln x -2 \ln (1-x)-I_0}{\epsilon}
\nonumber \\
&& \hspace*{-0.5cm}+ I_1-I_0 \ln x  +2 {\rm Li}_{2}(1-x) 
-\frac{1}{2} \ln^2 x +\ln^2(1-x)
- \frac{5 \pi^2}{6}+O(\epsilon) \Bigg\}  \; .
\end{eqnarray}
A useful relation in comparing these results with the real diagrams is
\begin{equation}
\Gamma(1+\epsilon)\Gamma(1-\epsilon)=1+\epsilon^2 \frac{\pi^2}{6}
+O(\epsilon^4)  \; .
\end{equation}
\section{Real integrals}
\setcounter{equation}{0}
In this appendix we will describe some of the integrals which occur 
in diagrams involving the emission of real partons.
As illustrated in Figs.~\ref{pd}(b,c),
we will denote the momenta of the emitted particles by $p_3$ and $p_4$
and the momentum of the `observed' parton line by $k$.
The phase space for one and two parton emission, 
and keeping $k^2$ and $n \cdot k$~fixed, is given by (we set $pn\equiv 1$
in the following)
\beqn
PS^{(1)}& =&
\int \frac{d^{d}p_3}{(2 \pi)^{d-1}} \delta^+(p_3^2)  
\nonumber \\
&& \int d^d k \; \delta (x- n \cdot k) \; 
\delta(|k^2|+(p-p_3)^2) \delta^d(p-p_3-k) 
\eeqn
\beqn \label{2partps}
PS^{(2)}& =&
\int \frac{d^{d}p_3}{(2 \pi)^{d-1}} \delta^+(p_3^2)  
\int \frac{d^{d}p_4}{(2 \pi)^{d-1}} \delta^+(p_4^2)   \nonumber \\
&& \int d^d k \; \delta (x- n \cdot k) \; 
\delta(|k^2|+(p-p_3-p_4)^2) \delta^d(p-p_3-p_4-k)  \; .
\nonumber \\
\eeqn
Integrating over irrelevant angles in $d$ dimensions we have 
for the transverse phase space,
\beq
\int d^{d-2}k_T= \frac{\pi^{\half-\ep}}{\Gamma(\half-\ep)}
\int d k_T^2 \; k_T^{-2 \ep} \int_0^{\pi} d\theta_1 \sin^{-2 \epsilon} 
\theta_1  \; .
\eeq
If the integrand is independent of $\theta_1$ we can integrate further
to obtain
\beq
\int d^{d-2}k_T= \frac{\pi^{1-\ep}}{\Gamma(1-\ep)}
\int_0^{|k^2| (1-x)} d k_T^2 \; k_T^{-2 \ep}  \; .
\eeq
\def\tita{Crossed ladder diagrams: topology b}
\subsection{\tita}
\begin{figure}[htb]
\vspace{8cm}
\includegraphics{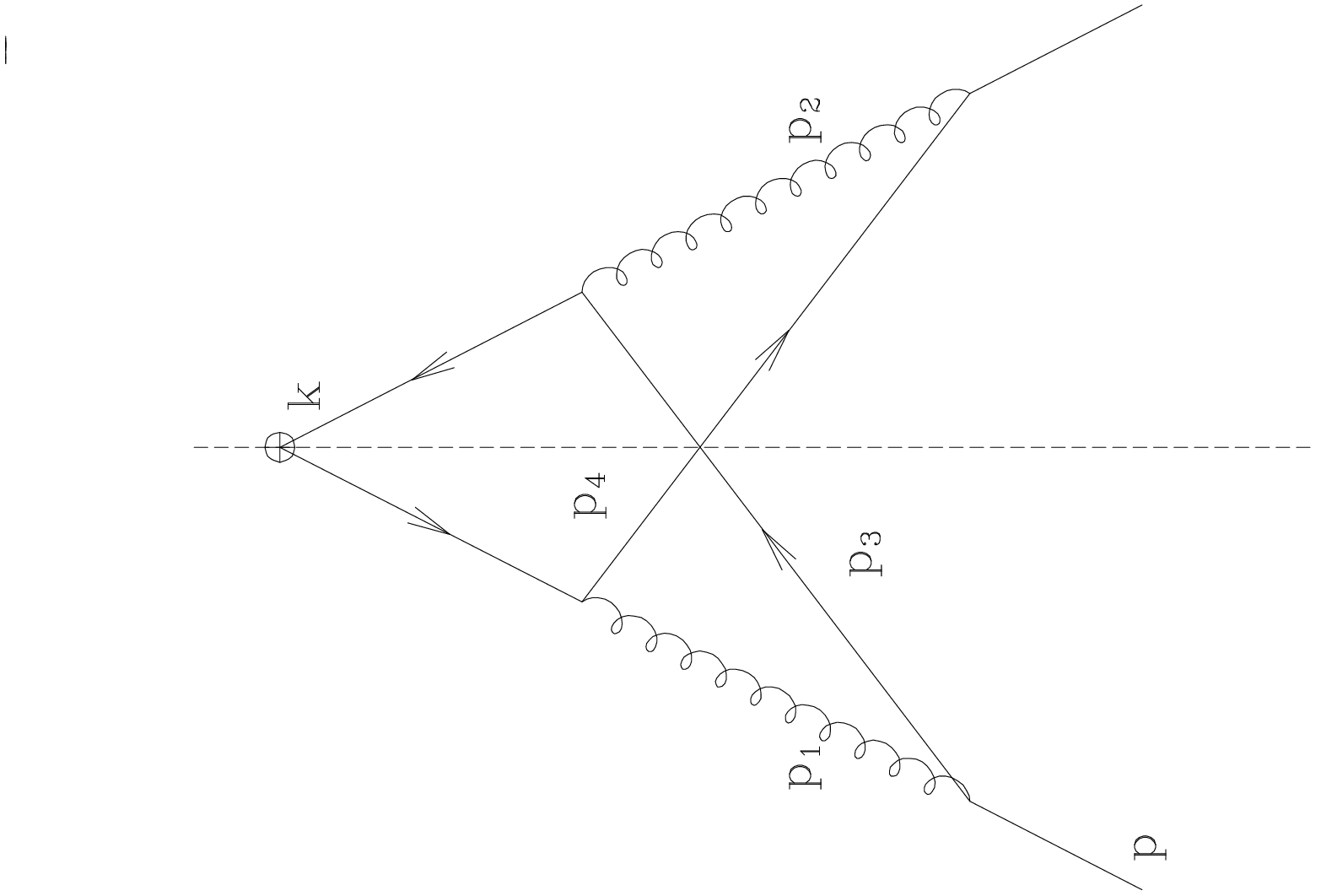}
\caption{An example of the graph of topology (b)}
\label{anomb}
\end{figure}

Here we shall describe the integrals needed for the evaluation of the 
crossed ladder diagram (topology b) as shown in Fig.~\ref{anomb}.
We introduce a notation for the real parton momenta such that,  
\beqn \label{crossedframe}
p_3 ^\mu &=& z_1 p^\mu + \frac{\tusq}{2 z_1 } n^\mu - \tu^\mu  \nonumber \\
p_4 ^\mu &=& z_2 p^\mu + \frac{\tdsq}{2 z_2 } n^\mu - \td^\mu 
\eeqn
with transverse momenta $\tu$, $\td$.
In terms of these variables the denominators which occur in the diagram
in Fig.~\ref{anomb} can be written as 
\beq
p_1^2=-\frac{\tu^2}{z_1},\;p_2^2=-\frac{\td^2}{z_2},\;
k^2=(p-p_3-p_4)^2=-a_1 \tu^2 -a_2 \td^2 -2 \tu \cdot \td
\eeq
where 
\beq
a_1=\frac{(1-z_2)}{z_1},\;
a_2=\frac{(1-z_1)}{z_2} \; .
\eeq

The general form of the matrix element which has to be integrated over
the phase space of Eq.~(\ref{2partps}) is
\beq \label{listofintegrands}
A(z_1,z_2)+B(z_1,z_2) \frac{\tu \cdot \td}{\tusq}
+C(z_1,z_2) \frac{\tu \cdot \td}{\tdsq}+D(z_1,z_2) 
\frac{(\tu \cdot \td)^2}{\tusq \tdsq}
\; .
\eeq
The integrations over $\tu$ and $\td$ are finite at small transverse momenta,
so that before the $k^2$ integration the expression is finite. 
Introducing the constants
\beqn
F &=& \frac{(4 \pi)^\ep }{16 \pi^2 \Gamma(1-\ep)} \nonumber \\
f &=& \frac{\pi^{1-\ep} }{\Gamma(1-\ep)}  \; 
\eeqn
we have in the frame specified by Eq.~(\ref{crossedframe}):
\beqn \label{framebphasespace}
PS^{(2)}&=&F^2 \; \int \frac{dz_1}{z_1}\; \frac{dz_2}{z_2} \; 
\delta(1-x-z_1-z_2)
\frac{d^{d-2} \tu}{f} \frac{d^{d-2} \td}{f} \nonumber \\
&& \delta \Bigg(|k^2| = a_1 \tu^2 +a_2 \td^2 +2 \tu \cdot \td \Bigg)  \; .
\eeqn
Since the integrals over the transverse momenta are finite we may 
take the limit
$d \rightarrow 4$. The values of the integrals for the integrands which occur
in Eq.~(\ref{listofintegrands}) and the phase space weight given in 
Eq.~(\ref{framebphasespace}) are collected in Table~\ref{table_realb}.
The remaining one dimensional integrals are easily performed.
\renewcommand\arraystretch{2.5}
\setcounter{table}{4}
\begin{table} 
\begin{center}
\begin{tabular}{|l|l|} 
\hline
Integrand & Integral in units of $F^2 \frac{|k^2|}{x}  \int dz_1 d z_2  \;
\delta(1-z_1-z_2-x)\theta(z_1)\theta(z_2)$\\
\hline
$<1>$ &$ 1 $\\
\hline
$<\frac{\tu.\td}{\tusq}>$ & 
$ -\frac{z_1}{1-z_2 } $\\
\hline
$<\frac{\tu.\td}{\tdsq}>$ & 
$ -\frac{z_2}{1-z_1 } $\\
\hline
$<\frac{(\tu.\td)^2}{\tusq \tdsq}>$ & 
$ \Big[1 
 + \half  \frac{x}{z_1 z_2 } \ln\left(\frac{x}{(1-z_1)(1-z_2)}\right) 
\Big] $\\
\hline
\end{tabular}
\label{table_realb}
\caption{Real integrals for crossed ladders}
\end{center}
\end{table}
\renewcommand\arraystretch{1.}
\subsection{Real diagrams: topology cd}
\begin{figure}[htb]
\vspace{8cm}
\includegraphics{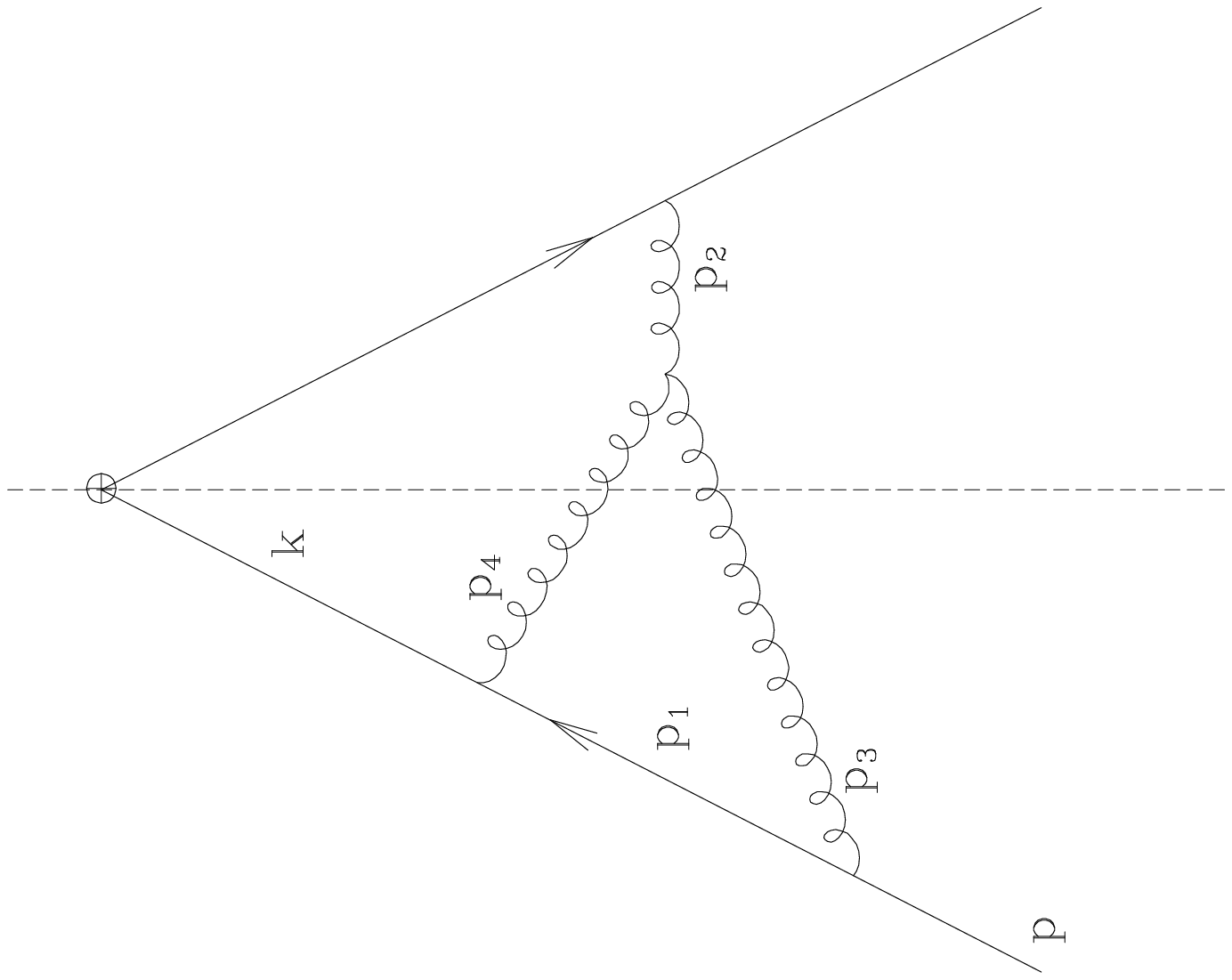}
\caption{An example of the graph of topology (cd)}
\label{anomcd}
\end{figure}
\renewcommand\arraystretch{2.5}
\setcounter{table}{5}
\begin{table} 
\begin{center}
\begin{tabular}{|l|l|} 
\hline
Integrand & Value of integral in units of $F^2 |k^2|^{-2\ep} 
\frac{\Gamma^2(1-\ep)}{\Gamma(1-2 \ep)}$ \\
\hline
$<1>$ &$ \frac{|k^2|}{1-2\ep}
\int_x^1 \frac{dz}{x} \bigg( \frac{x}{(1-z)(z-x)} \bigg)^\ep $ \\
\hline
$<\frac{1}{|p_1^2|}>$ & $-\frac{1}{\ep} 
\; \int_x^1 \frac{dz}{z} \bigg(\frac{x}{(1-z)(z-x)} \bigg)^\ep$ \\
\hline
$<\frac{1}{|p_2^2|}> $ & $ -\frac{1}{\ep} 
\; \int_x^1 \frac{dz}{(1-x)} \; \bigg( \frac{x}{(1-z)(z-x)} \bigg)^\ep $ \\
\hline
$<\frac{1}{|p_1^2| |p_2^2|}>$ & 
$ -\frac{2}{\ep} \frac{1}{|k^2|}  \int_x^1 \frac{dz}{(1-z)^{1+2\ep}} 
 \bigg(1+ \ep \ln\big(\frac{z(1-x)}{z-x}\big)
 +\ep^2 \frac{\pi^2}{6} +O \left( \ep^2 (1-z) \right) \bigg)$ \\
\hline
\end{tabular}
\caption{Real integrals}
\label{table_real}
\end{center}
\end{table}
\renewcommand\arraystretch{1.}
The calculation of the topology (b) graphs was facilitated by the fact that 
before integration over $k^2$ the kernel was finite. The situation is more 
complicated for topology (cd), which has two cuts involving either one or 
two real partons. The cut with two real partons (gluons) is shown in 
Fig.~\ref{anomcd} which also shows the definition of the kinematics.

We perform a light-cone decomposition of the light-like momenta 
$p_3$ and $p_4$:
\beqn
p_3 ^\mu &=& (1-z) p^\mu + \frac{\tusq}{2(1-z)} n^\mu - \tu^\mu  
\nonumber \\
p_4 ^\mu &=& z (1-y) p^\mu + \frac{\tdsq}{2 z (1-y)} n^\mu -\td^\mu 
\eeqn
where $zy=x$. It is expedient to perform a change of variables,
\beqn
\tu &\rightarrow& \sqrt{\frac{1-z}{1-y}} 
\ru \nonumber \\
\td &\rightarrow& \sqrt{\frac{1-y}{1-z}} \big(\rd   -(1-z) \ru \big) \; , 
\eeqn 
so that $p_3$ and $p_4$ become
\beqn
p_3 ^\mu &=& (1-z) p^\mu + \frac{\rusq}{2(1-y)} n^\mu 
- \sqrt{\frac{1-z}{1-y}} \ru^\mu \\
p_4 ^\mu &=& z (1-y) p^\mu 
+ \frac{\left( \rd-(1-z)\ru \right)^2}{2 z (1-z)} n^\mu 
-\sqrt{\frac{1-y}{1-z}} \left( \rd^\mu-(1-z) \ru^\mu \right) \; . \nonumber  
\eeqn
With this choice the propagators of the diagram in Fig.~\ref{anomcd} 
can be written as
\beqn
|k^2| &= -(p-p_3-p_4)^2 &= 
\frac{\rusq y }{(1-y)}+\frac{\rdsq }{(1-z)} \\
|p_1^2|&=-(p-p_3)^2   &= \frac{\rusq}{1-y}\\
\label{p2sqdef}
|p_2^2|&=(p_3+p_4)^2  &= \frac{\rusq +\rdsq -2 \ru \cdot \rd}{z} \\
n\cdot p_3 = 1-z, &&  n \cdot p_4 = z (1-y) \; .
\eeqn
Hence we have that in this frame,
\beqn \label{2partpsmod}
PS^{(2)}&=&F^2 \; \int \frac{dz}{(1-z)}\; \frac{dy}{(1-y)} \; \delta(x-y z)
\frac{d^{d-2} \ru}{f} \frac{d^{d-2} \rd}{f} \nonumber \\
&& \delta \Bigg(|k^2|=
\frac{\rusq y }{(1-y)}+\frac{\rdsq }{(1-z)}
\Bigg) \; .
\eeqn
If we are integrating over quantities which do not depend on angles we 
may write Eq.~(\ref{2partpsmod}) in the form
\beq  \label{ps2new}
PS^{(2)}=F^2 \; |k^2|^{1-2 \ep}
\int_x^1 \frac{dz}{x} \; \Bigg(\frac{x}{(1-z)(z-x)} \Bigg)^\ep
\; \int_0^1 d \omega \; \omega^{-\ep} (1-\omega)^{-\ep}
\eeq
where the rescaled transverse momentum $\omega$ is defined as
\beq 
|p_1^2| =\frac{\rusq}{(1-y)}= \frac{\omega z |k^2|}{x}  \; .
\eeq
The results for the integrals are given in Table~\ref{table_real}.

We note at this point that Eq.~(\ref{ps2new}) is also suitable for dealing
with a light-cone gauge denominator term like $1/(n\cdot p_3)$ in the 
matrix element. For instance, we obtain
\beq
<\frac{1}{n\cdot p_3}> = F^2 |k^2|^{1-2 \epsilon} \frac{1}{x}
\int_x^1 dz \mbox{PV} \Bigg( \frac{1}{1-z} \Bigg)
=F^2 |k^2|^{1-2 \epsilon} \frac{1}{x} \left( I_0+\ln (1-x) \right)
\eeq
where $I_0$ is as defined in Eq.~(\ref{i0def}). 

If we have a denominator which depends on the angle between $\ru$ and $\rd$, 
the integral is more complicated. For example when we have the denominator 
as given by Eq.~(\ref{p2sqdef}) the angular integration 
splits into two regions, $\rusq>\rdsq$ and $\rdsq>\rusq$:
\beq \label{onep2den}
<\frac{1}{|p_2^2|}> = F^2 \; |k^2|^{-2\ep}
\; \int_x^1 \frac{dz}{(1-z)} \; \frac{z}{x} \; 
\Bigg( \frac{x}{(1-z)(z-x)} \Bigg)^\ep I(\alpha)
\eeq
\beqn
I(\alpha)&=& \Bigg\{
\int_0^{\frac{1}{1+\alpha}} 
d \omega \; \omega^{-\ep} (1-\omega)^{-1-\ep} 
{}_2 F_1 (1,1+\ep;1-\ep;\frac{\alpha \omega}{1-\omega}) \nonumber \\
&&+\frac{1}{\alpha}
\int_{\frac{1}{1+\alpha}}^1 d \omega \; \omega^{-1-\ep} (1-\omega)^{-\ep} 
{}_2 F_1 (1,1+\ep;1-\ep;\frac{1-\omega}{\alpha \omega})  \Bigg\}
\eeqn
where 
\beq \label{alphadefinition}
\alpha=\frac{(z-x)}{x(1-z)}  \; .
\eeq
Now by redefinition of variables the two integrals give
\beq
I(\alpha)= J(\alpha)+\frac{1}{\alpha}J(\frac{1}{\alpha})
\eeq
where
\beqn
J(\alpha)
&=& \alpha^{-\ep} \int_0^1 dv v^{-\ep} 
(1+\alpha v)^{-1+2 \ep} {}_2 F_1  (1,1+\ep; 1-\ep;v) \nonumber \\ 
&&\equiv -\frac{1}{2 \ep} \alpha^{-\ep} (1+\alpha)^{-1+2\ep} 
{}_2 F_1 (1,-2 \ep; 1-\ep;\frac{\alpha}{1+\alpha})   \; .
\eeqn
Hence combining using the identity,
\beq
{}_2 F_1 (1,-2 \ep; 1-\ep;z) +
{}_2 F_1 (1,-2 \ep; 1-\ep;1-z)  =
2 \; \frac{\Gamma^2(1-\ep)}{\Gamma(1-2 \ep)} z^\ep (1-z)^\ep
\eeq
one obtains
\beq
I(\alpha)=
-\frac{1}{\ep} \frac{\Gamma^2(1-\ep)}{\Gamma(1-2 \ep)} \frac{1}{1+\alpha}
\; .
\eeq
Thus the final result is as given in Table~\ref{table_real}.
This result can be obtained much more easily by performing a shift 
of the transverse momenta so that $|p_2^2|$ only depends on a single 
transverse momentum.

If we now add a second denominator such a shift is no longer useful. The 
scalar integral with two denominators is given by a simple modification of 
Eq.~(\ref{onep2den}),
\beqn
<\frac{1}{|p_1^2| |p_2^2|}> &=& F^2 \; |k^2|^{-1-2\ep}
\; \int_x^1 \frac{dz}{(1-z)^{1+2 \ep}} \; 
\nonumber \\
&&\Bigg\{ 
\alpha^{-\ep}
\int_0^{\frac{1}{1+\alpha}}  d \omega \; \omega^{-1-\ep} (1-\omega)^{-1-\ep} 
{}_2 F_1 (1,1+\ep;1-\ep;\frac{\alpha \omega}{1-\omega})
\nonumber \\
&&+\alpha^{-1-\ep}
\int_{\frac{1}{1+\alpha}}^1 d \omega \; \omega^{-2-\ep} (1-\omega)^{-\ep} 
{}_2 F_1 (1,1+\ep;1-\ep;\frac{1-\omega}{\alpha \omega}) \Bigg\}
\nonumber \\
\eeqn
where $\alpha$ is as given in Eq.~(\ref{alphadefinition}). 
By change of variables this integral may be 
further written as
\beqn
<\frac{1}{|p_1^2| |p_2^2|}> &=& F^2 \; |k^2|^{-1-2\ep}
\; \int_x^1 \frac{dz}{(1-z)^{1+2 \ep}}  K(\alpha)
\eeqn
where 
\beqn
K(\alpha)&=&\Bigg\{ 
\int_0^{1}  d v \; v^{-(1+\ep)} \; \Big ( \frac{v}{\alpha}+1 \Big)^{2\ep} 
\; {}_2 F_1 (1,1+\ep;1-\ep;v)
\nonumber \\
&&+ \int_{0}^1 d v \; v^{-\ep} \; \Big(\frac{1}{\alpha}+v \Big)^{2\ep} 
\; {}_2 F_1 (1,1+\ep;1-\ep;v)
\Bigg\}   \; .
\eeqn
The partial result for $K(\alpha)$ which is sufficient for our purposes is
\beq
K(\alpha)=-\frac{2}{\ep}  
\frac{\Gamma^2(1-\ep)}{\Gamma(1-2 \ep)}  
\bigg(1+ \ep \ln\left( \frac{z(1-x)}{z-x} \right)
+\ep^2 \frac{\pi^2}{6} +O \left( \ep^2 (1-z)\right) \bigg)  \; .
\eeq
So the final result for the integral is as given in Table~\ref{table_real}.
\begin{reflist}
\item \label{Leib}
G. Leibbrandt, \pr{D29}{1699}{84}; \\
For an overview see also: G. Leibbrandt, \rmp{59}{1067}{87}.
\item \label{AT}
A. Andrasi and J. C. Taylor, \np{B310}{222}{88}.
\item \label{CFP}
G. Curci, W. Furmanski and R. Petronzio, \np{B175}{27}{80}.
\item \label{FP}
W. Furmanski and R. Petronzio, \pl{97B}{437}{80}.
\item \label{Mandel}
S. Mandelstam,  \np{B213}{149}{83}.
\item \label{Bass}
A. Bassetto, \pr{D47}{727}{93}.
\item \label{Vogelsang}
W. Vogelsang, Rutherford preprint, RAL-TR-95-071, hep-ph/9512218.
\item \label{HamVN}
R. Hamberg, PhD Thesis, University of Leiden, 1991; \newline
R. Hamberg and W. Van Neerven, \np{B379}{143}{92}.
\item \label{ColSc}
J.C. Collins and R.J. Scalise, \pr{D50}{4117}{94}.
\item \label{EGMPR}
R.K. Ellis \etal, \np{B152}{285}{79}.
\item \label{Larin}
S.A. Larin \etal, \np{B427}{41}{94}.
\item \label{AlPa}
G. Altarelli and G. Parisi, \np{B126}{298}{77}.
\item \label{GLAP}
Yu.~L.~Dokshitzer \JETP{46}{641}{77}; \newline
\cf~L.~N.~ Lipatov, \sj{20}{95}{75}; \newline
V.N.~Gribov and L.N.~ Lipatov, \sj{15}{438}{72}.
\item \label{HeWa}
R.T. Herrod and S. Wada, \pl{96B}{195}{80}.
\end{reflist}
\end{document}

%% file: mathmac.tex
\newskip\humongous \humongous=0pt plus 1000pt minus 1000pt

\newif\ifdtup

	\def\cf{\hbox{\it cf.}{}}
\def\etal{\hbox{\it et al.}}    

\def\beq{\begin{equation}}
\def\eeq{\end{equation}}

\def\beqn{\begin{eqnarray}}
\def\eeqn{\end{eqnarray}}
\relax

\def\dotx{\dotx{\dot\overline{x}}}

\relax

%% file: webmac.tex
\def\cf{{\it cf.\ }}

\def\etal{{\it et al.}}

\def\half{\mbox{\small $\frac{1}{2}$}}

\newcommand{\slsh}{\rlap{$\;\!\!\not$}}     
\long\def\query#1{%
\hskip 0pt{\vadjust{\everypar={}\small\vtop to 0pt{\hbox{}%
\vskip -13pt\rlap{\hbox to 38.5pc{\hfil{\vtop{\hsize=8pc\tolerance=6000%
\hfuzz=.5pc\rightskip=0pt plus 3em\noindent#1}}}}\vss}}}%
}

%% file: refdef.tex
\def\reflist{\section*{References
\markright{REFERENCES}
}\footnotesize\list
        {\footnotesize\arabic{enumi})\hfill}{\settowidth\labelwidth{[999]}
        \leftmargin\labelwidth
        \advance\leftmargin\labelsep\usecounter{enumi}}}

\relax
\relax
\def\pl#1#2#3{
        {\it Phys.\ Lett.\ }{\bf #1} (19#3) #2}

\def\rmp#1#2#3{
        {\it Rev.\ Mod.\ Phys.\ }{\bf #1} (19#3) #2}

\def\pr#1#2#3{
        {\it Phys.\ Rev.\ }{\bf #1} (19#3) #2}
\def\np#1#2#3{
        {\it Nucl.\ Phys.\ }{\bf #1} (19#3) #2}

\def\nc#1#2#3{
        {\it Nuovo Cimento } {\bf #1} (19#3) #2}
\def\JETP#1#2#3{
        {\it Sov.\ Phys.\ JETP} {\bf #1} (19#3) #2}
\def\sj#1#2#3{
        {\it Sov.\ J.\ Nucl.\ Phys.\ } {\bf #1} (19#3) #2}

\relax

%% file: command.tex
\def\half{\mbox{\small $\frac{1}{2}$}}
\def\tu{{\bf t_1}}
\def\ep{\epsilon}
\def\td{{\bf t_2}}
\def\ru{{\bf r_1}}
\def\rd{{\bf r_2}}
\def\tusq{{\bf t_1^2}}
\def\tdsq{{\bf t_2^2}}
\def\rusq{{\bf r_1^2}}
\def\rdsq{{\bf r_2^2}}
\def\ason2pi{\frac{\alpha_s}{2 \pi}}
\def\nc{N_C}
\def\as{\alpha_s}

\def\lm{\lambda_-}

\def\lp{l_+}
\def\kp{k_+}
\def\pp{p_+}
\def\ppp{p^\prime_+}
\def\lm{l_-}